\documentclass{jfm}
\usepackage{graphicx}
\usepackage{amsmath}
\usepackage{amssymb}
\usepackage{xcolor}
\usepackage{epstopdf, epsfig}
\usepackage{multirow}
\usepackage{booktabs}
\usepackage{hyperref}
\usepackage{longtable}

\shorttitle{A Diffuse-Interface Marangoni Instability}
\shortauthor{X. Li., D. Wan, H. Hao, C. Diddens, M. Zhang, and H. Tan}

\title{A Diffuse-Interface Marangoni Instability}

\author{Xiangwei Li\aff{1},
  Dongdong Wan\aff{2},
  Haohao Hao\aff{1},
  Christian Diddens\aff{3},
  Mengqi Zhang\aff{2},
  \and Huanshu Tan\aff{1}
  \corresp{\email{tanhs@sustech.edu.cn}}
}

\affiliation{\aff{1}Multicomponent Fluids group, Center for Complex Flows and Soft Matter Research, Department of Mechanics and Aerospace Engineering, Southern University of Science and Technology, Shenzhen, 518055, Guangdong, P.R. China
\aff{2}Department of Mechanical Engineering, National University of Singapore, 9
Engineering Drive 1, 117575, Republic of Singapore
\aff{3}Physics of Fluids Department, Max-Planck Center Twente for Complex Fluid Dynamics and J. M. Burgers Centre for Fluid Dynamics, University of Twente, P. O. Box 217, 7500AE Enschede, The Netherlands
}

\begin{document}

\maketitle

\begin{abstract}
We investigate a novel Marangoni-induced instability that arises exclusively in diffuse fluid interfaces, absent in classical sharp-interface models. 
Using a validated phase-field Navier–Stokes–Allen–Cahn framework, we linearize the governing equations to analyze the onset and development of interfacial instability driven by solute-induced surface tension gradients. 
A critical interfacial thickness scaling inversely with the Marangoni number, $\delta_\mathrm{cr} \sim Ma^{-1}$, emerges from the balance between advective and diffusive transport. 
Unlike sharp-interface scenarios where matched viscosity and diffusivity stabilize the interface, finite thickness induces asymmetric solute distributions and tangential velocity shifts that destabilize the system. 
We identify universal power-law scalings of velocity and concentration offsets with a modified Marangoni number $Ma^\delta$, independent of capillary number and interfacial mobility. 
A critical crossover at $Ma^\delta \approx 590$ distinguishes diffusion-dominated stabilization from advection-driven destabilization. 
These findings highlight the importance of diffuse-interface effects in multiphase flows, with implications for miscible fluids, soft matter, and microfluidics where interfacial thickness and coupled transport phenomena are non-negligible.
\end{abstract}


\section{Introduction}
\label{sec:1}

Marangoni instability, arising from the transverse mass transfer of surface-active species across a fluid interface~\citep{sternling_interfacial_1959,Kovalchuk_2006,schwarzenberger_pattern_2014}, plays a fundamental role in multicomponent flow systems, such as evaporating droplets~\citep{tan2016evaporation,lohse2020physicochemical,diddens2024non}, self-motived  droplets~\citep{levich1969surface,michelin2023self}, microextraction solutions~\citep{eckert_chemo-marangoni_2012,mitra2020efficient}, and liquid films~\citep{darhuber2003marangoni,sultan2005evaporation}.
Its onset and evolution are governed by fluid property contrasts, particularly viscosity and mass diffusivity ratios between the interacting layers.
Traditional linear stability analysis, based on an idealized sharp and non-deformable interface, establishes critical instability criteria~\citep{sternling_interfacial_1959,reichenbach_linear_1981}.
However, direct numerical simulations with a finite-thickness, deformable interface validate these criteria while also revealing that excessive interface thickness can itself induce Marangoni instability~\citep{verschuerenDiffuseinterfaceModellingThermocapillary2001,LI2025105073}.
This study seeks to further clarify the role of interface thickness in Marangoni instability and the underlying mechanisms.

\begin{figure}
  \centerline{\includegraphics[width=0.85\linewidth]{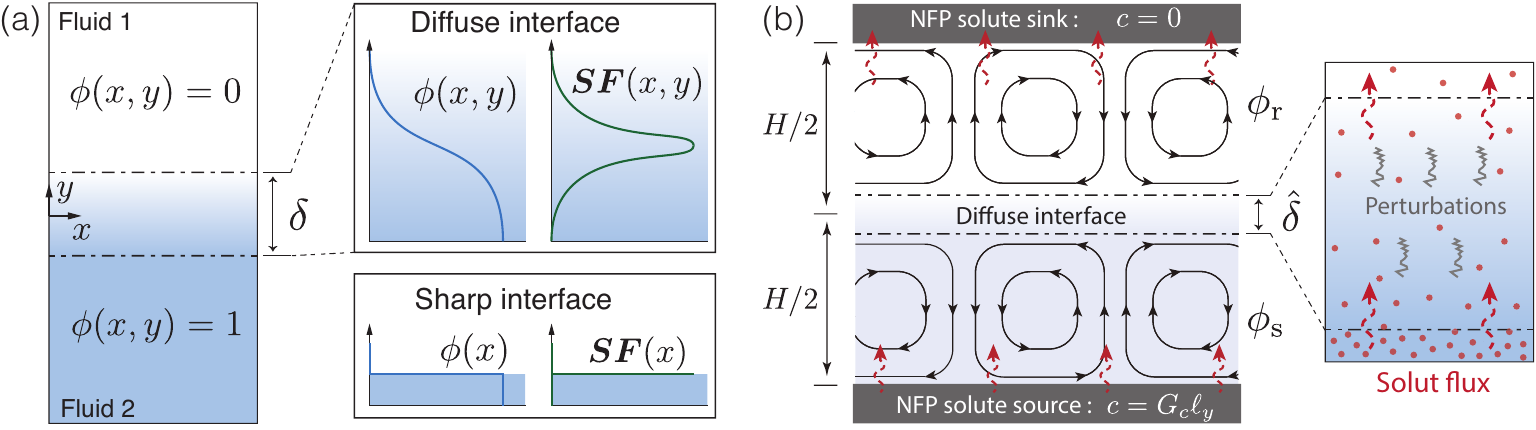}}
  \caption{
  (a) Schematic of a two-liquid layer system with a constant gradient of solute $\nabla c$ in $y$-direction. 
  (b) Phase-field variable $\phi_{eq}$ distribution in the diffuse-interface method.}
  \label{fig:1}
\end{figure}

A perfectly sharp fluid interface -- one with zero thickness -- does not exist.
Even in immiscible liquids, the phase transition region spans nanometer scales~\citep{mitrinovic2000noncapillary,senapati2001computer}.
Jump conditions are commonly employed to approximate interface by neglecting the continuous variation of fluid properties across it~\citep{scardovelli1999direct,rederViscousStressApproximations2024}, provided that the characteristic system size is much larger than the interfacial thickness $\hat{\delta}$.
However, this assumption breaks down when the interfacial thickness becomes comparable to the length scale of the phenomena being examined~\citep{anderson1998diffuse} -- either due to interface thickening in miscible or near-critical fluids, or the shrinking of the system in sub-micron liquid systems~\citep{cahn1958free,stanley1987introduction,joseph1993fluid}.
In such cases, the transition occurs over a finite region still large compared to the molecular scale, motivating the diffuse-interface concept originally introduced by~\cite{rayleigh1892xx} and~\cite{van1893verhandel}, as illustrated in Figure~\ref{fig:1}a.
In such cases, the variation of properties within the interface can no longer be ignored~\citep{cahn1958free}.

Interfacial surface tension arises from tangential forces across the finite-thickness transition layer between two fluids, where anisotropic intermolecular forces act parallel to the interface~\citep{berry1971molecular,marchand2011surface}. 
Surface-active solutes locally reduce tension according to a first-order approximation, $\gamma=\gamma_0-\beta c$, where $c$ is the solute concentration and $\beta$ is a proportionality constant.
While classical theories assume a sharp interface, this concept extends to finite-thickness interfaces, where surface force $\boldsymbol{SF}$ may develop heterogeneously across sublayers (Fig.~\ref{fig:1}a inset), driving localized solutal Marangoni flows~\citep{anderson1998diffuse,hao2025enhanced}.
These flows typically form symmetric convection rolls centered at stagnation points (Fig.~\ref{fig:1}b), and their multi-scale interactions lead to interfacial turbulence~\citep{ruckenstein1964occurrence}. 
The linear theory by~\cite{sternling_interfacial_1959}, based on the sharp-interface model, highlights the roles of viscosity and diffusivity ratios between source ($\phi_\text{s}$) and receiver ($\phi_\text{r}$) fluids, but is intrinsically incapable of capturing the coupled process~\citep{manikantanSurfactantDynamicsHidden2020} within diffuse interfaces (Fig.~\ref{fig:1}b inset).
The interfacial thickness impact on the onset and development of Marangoni instability warrants further study.

Phase-field methods model fluid interfaces with finite thickness and have been successfully applied to both diffuse-interface problems and sharp-interface two-phase Navier-Stokes flows~\citep{anderson1998diffuse,jacqmin_calculation_1999}.
Derived from fluid free energy formulations~\citep{cahn1958free}, the phase field recasts the moving-boundary problem into a continuous framework.
In simulations, interface distortion due to flow is regulated by the mobility parameter $M_{\phi}$, balancing anti-diffusion and convection of the phase field $\phi$~\citep{demont_numerical_2023}.
As the interface thickness $\hat{\delta}$ decreases, phase-field solutions converge toward sharp-interface results~\citep{magalettiSharpinterfaceLimitCahn2013}.
The physic-based formulation and its asymptotic consistency make the phase-field method well-suited for capturing interfacial flow dynamics with finite-thickness effects.

The paper is organized as follows. Section~\ref{sec:2} introduces the problem formulation using a phase-field model and its linearization. Section~\ref{sec:3} identifies the instability unique to diffuse interfaces, establishes a scaling law for the critical thickness, and reveals the underlying mechanisms through interface offset analysis. Section~\ref{sec:4} concludes with key findings and their physical implications.

\section{Problem Formulation and Linearization}
\label{sec:2}

We use a validated phase-field Navier–Stokes–Allen–Cahn (NSAC) model to study solute-driven interfacial instability~\citep{LI2025105073}. A brief description follows.

As depicted in Figure~\ref{fig:1}b, the system is characterized by a solute concentration gradient $G_c$ oriented normal to a diffuse interface separating two fluid phases.
These fluids are denoted by $\phi_\text{s}$ and $\phi_\text{r}$, corresponding to the solute-supplying and solute-receiving phases, respectively. 
Both phases are confined between two flat, no-flow-penetration (NFP) boundaries, while periodic boundary conditions apply laterally. 
The characteristic length is defined as one quarter of the spacing distance, i.e., $L=H/4$.
The interface is captured using the conservative Allen-Cahn equation in dimensionless form, which accounts for advective transport,
\begin{equation}
    \label{eq:AC}
        \frac{\partial \phi_{j}}{\partial t} + \nabla \cdot \left(\phi_{j} \boldsymbol{u} \right) = \frac{1}{{Pe}_\phi} \left\{ \nabla \cdot \left[ \nabla \phi_{j} - B \phi_{j} \left(1-\phi_{j}\right) \frac{\nabla \phi_{j}}{\left|\nabla \phi_{j}\right|} \right] + \alpha_j \right\},
\end{equation}
where $j=\text{s}, \text{r}$~(no $j$ summation), and $B=4/\delta \operatorname{artanh}(1-2\delta_{n})$ controls the dimensionless interfacial thickness $\delta = \hat{\delta}/L$, and $\delta_n$ defines the numerically resolvable transition width.
The equilibrium phase-field profile takes the form $\phi_{\mathrm{eq}}=\{1-\tanh [(y-2L)B/2]\}/2$ (Fig.~\ref{fig:1}a inset).
The Lagrange multipliers $\alpha_j$ enforce the constraint of $\phi_\text{s}+\phi_\text{r}=1$.

The governing equations for fluid momentum and solute transport in dimensionless form are given by, 
\begin{subequations}
    \label{eq:NS}
    \begin{gather}
        \frac{\partial \boldsymbol{u}}{\partial t} + \boldsymbol{u} \cdot \nabla \boldsymbol{u} = \frac{Sc}{Ma} \left\{ -\nabla p + \nabla \cdot \left[ \eta \left( \nabla \boldsymbol{u} + (\nabla \boldsymbol{u})^{T} \right) \right] + \frac{\boldsymbol{S F}}{Ca} \right\}, \quad 0  = \nabla \cdot \boldsymbol{u} , \label{eq:u}\\
      \frac{\partial c}{\partial t} = \frac{1}{Ma} \nabla \cdot \left[D (\nabla c)\right] - \nabla \cdot (c \boldsymbol{u}), \label{eq:c}
    \end{gather}
\end{subequations}
where the surface force term $\boldsymbol{SF} = \frac{1}{2} \left[ \gamma (\kappa_\text{s} \boldsymbol{n}_\text{s}+ \kappa_\text{r} \boldsymbol{n}_\text{r}) +  (\boldsymbol{I}_\text{s}-\boldsymbol{n}_\text{s}\boldsymbol{n}_\text{s} +\boldsymbol{I}_\text{r}-\boldsymbol{n}_\text{r}\boldsymbol{n}_\text{r})\nabla \gamma \right] W$ follows the formulation by~\cite{kim_phase-field_2012},
with $\boldsymbol{n}_j=\nabla \phi_j/|\nabla \phi_j|$, $\kappa_j=-\nabla \cdot \boldsymbol{n}_j$, and $(\boldsymbol{I}_j-\boldsymbol{n}_j\boldsymbol{n}_j) \cdot \nabla$ representing the surface gradient operator in terms of phase field $\phi_j$.
The surface tension decreases linearly with solute concentration under dilute conditions~\citep{Picardo_2016}, modeled as $\gamma=1-Ca~ c$, where $Ca$ is the capillary number.
The interface localization function is $W=A \phi_\text{s} \phi_\text{r} |\nabla\phi_\text{s}| |\nabla\phi_\text{r}|$, with $A=30/B$. 
Viscosity and diffusivity are set as $\eta =  \phi_{\text{s}} + \phi_{\text{r}}/\zeta_\eta$ and $D =  \phi_{\text{s}} + \phi_{\text{r}}/\zeta_D$.
We choose $\zeta_\eta=\zeta_D=1$ to isolate Marangoni instability driven by property ratios~\citep{sternling_interfacial_1959}.
Gravity is neglected and density ratio is unity.

The dimensionless groups appearing in the equations are Schmidt number $Sc=\eta/(\rho D)$, capillary number $Ca = U\eta/\gamma_0$, and Marangoni number $Ma=UL/D$.
The characteristic velocity is defined as the Marangoni flow $U=\beta |G_c| L/\eta$ that prevails in the stationary flow, with $G_c=\Delta c/(4L)$, $\beta=\text{d} \gamma/\text{d} c$.
The Péclet number for the phase field is given by $Pe_\phi=UL/M_{\phi}$, where the interfacial mobility $M_\phi \propto \delta^{\alpha}$, with an optimal scaling parameter of $\alpha \approx 1.7$, describes the rate of convergence of the diffuse-interface solution to the sharp-interface solution~\citep{demont_numerical_2023}.

The governing equations (Eqs.~\ref{eq:AC}, \ref{eq:NS}) are linearized around a base state $\boldsymbol{Q}^b = (\boldsymbol{U}^b, P^b, C^b, \phi_\text{s}^b)^T$ with zero velocity $\boldsymbol{U}^b=\boldsymbol{0}$, uniform pressure $P^b=\text{const}$, an equilibrium phase field $\phi_\text{s}^b=\phi_{\text{eq}}$, and a steady solute gradient $C^b = G_c(H - y)$.
The perturbations are assumed to take the form $\boldsymbol{q}' = \tilde{\boldsymbol{q}}(y) \text{e}^{\text{i}kx-\text{i}\omega t} + \text{c.c.}$, where $k$ is the wavenumber, and
$\omega=\omega_{r} + \text{i} \omega_i$ is the complex frequency. 
A positive growth rate ($\omega_i>0$) indicates linear instability, while $\omega_i<0$ implies stability. 
The real part $\omega_r$ represents the oscillation frequency. 
This leads to a generalized eigenvalue problem of the form, $  \omega \widetilde{\boldsymbol{M}} \tilde{\boldsymbol{q}}=  \widetilde{\boldsymbol{L}} \tilde{\boldsymbol{q}}$, where $\widetilde{\boldsymbol{M}}$ and $\widetilde{\boldsymbol{L}}$ are linear operators derived from the linearized equations as following,
\begin{subequations}\label{eq:linear_eq_Fourier}
  \begin{gather}
  -\text{i}\omega{ \tilde{\phi} _\text{s}}= -\frac{\mathrm{d} \phi_{ \text{s}}^b}{\mathrm{d} y }\tilde{v} + \frac{1}{Pe_\phi} \frac{\mathrm{d}^2}{\mathrm{d} y^{2}} \tilde{\phi}_{ \text{s}} -\frac{B}{Pe_\phi}   \left[(2\phi_{ \text{s}}^b -1)\frac{\partial \tilde{\phi} _\text{s}}{\partial y } + 2\frac{\mathrm{d} \phi_{ \text{s}}^b}{\mathrm{d} y } \tilde{\phi} _\text{s}\right],\label{eq:linear_phi}\\
  -\text{i}\omega{ \tilde{c}} = -\frac{\mathrm{d} C^b}{\mathrm{d} y }\tilde{v} + \frac{1}{Ma} (\frac{\mathrm{d}^2}{\mathrm{d} y^{2}} - k^2)\tilde{c},\label{eq:linear_c}\\
  -\text{i}\omega\tilde{u} = \frac{Sc}{Ma} \left[-\text{i}k\tilde{p} + (\frac{\mathrm{d}^2}{\mathrm{d} y^{2}} - k^2)\tilde{u}  +\frac{W^b}{Ca} \frac{1}{\left|\nabla\phi^b_\text{s}\right|}\frac{\mathrm{d} \Gamma^b}{dy} \text{i} k {\tilde{\phi}_\text{s}}-\text{i}k{\tilde{c}} W^b
 \right], \label{eq:linear_u}\\
  -\text{i}\omega{\tilde{v}} = \frac{Sc}{Ma} \left[-\frac{\partial \tilde{p}}{\partial y } + (\frac{\mathrm{d}^2}{\mathrm{d} y^{2}} - k^2) \tilde{v} +\frac{W^b}{Ca} \frac{1}{\left|\nabla\phi^b_\text{s}\right|}\Gamma^b(-k^2){\tilde{\phi} _\text{s}}  \right],\label{eq:linear_v}\\
  0=\text{i}k{ \tilde{u}} + \frac{\partial \tilde{v}}{\partial y },
  \end{gather}
\end{subequations}
where the base-state of surface tension $\Gamma^b=1- Ca C^b$ and weight function $W^b=A \phi_{\mathrm{s}}^b(1-\phi_{\mathrm{s}}^b)\left|\nabla  \phi_{\mathrm{s}}^b\right|^2$.
We refer the reader to~\cite{LI2025105073} for the detailed derivation.

\section{Results and Discussions}
\label{sec:3}
\subsection{Unveiling instabilities in diffuse interfaces}
\label{subsec:3.1}

\begin{figure}
  \centerline{\includegraphics[width=0.8\linewidth]{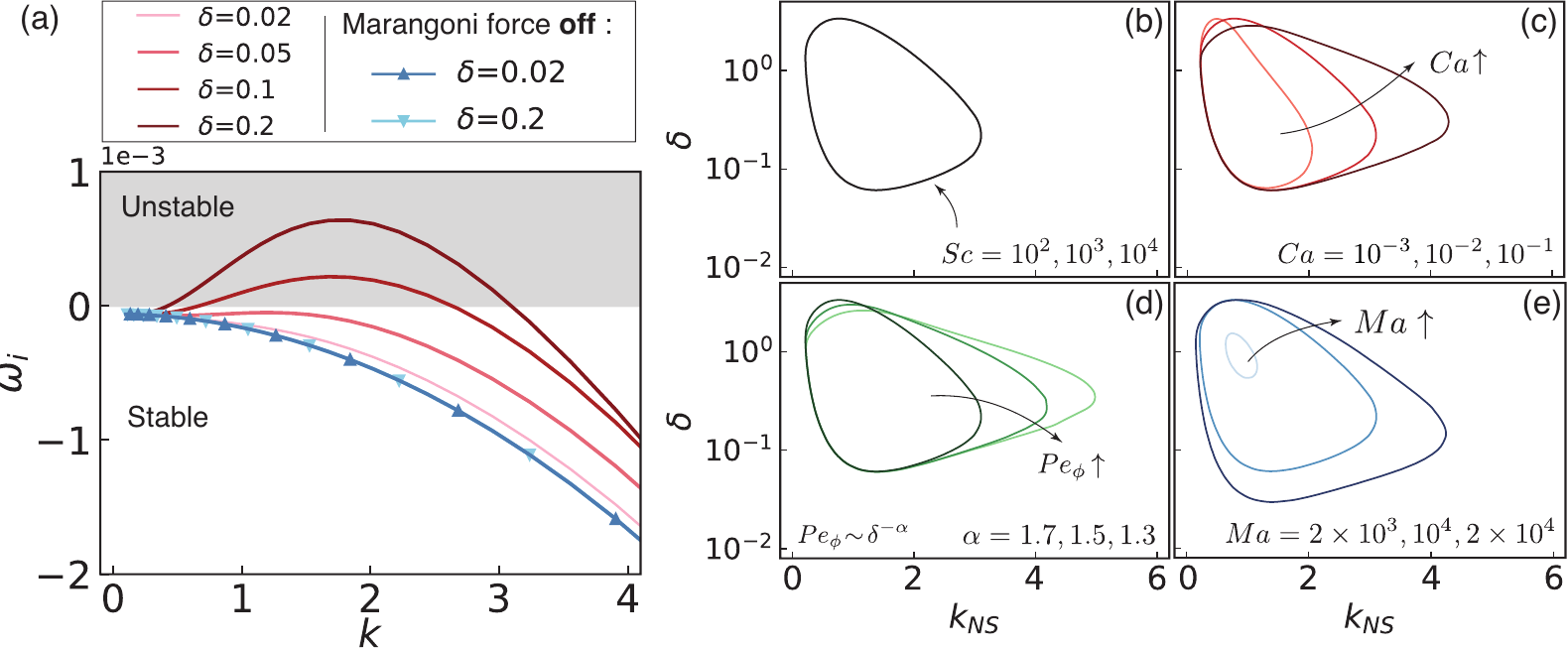}}
\caption{(a) Growth rate $\omega_i$ versus wavenumber $k$ for different $\delta$. The gray-shaded area indicates $\omega_i > 0$, i.e., instability. Stability is restored when Marangoni force is off. Parameters: $Ma=10^4$, $Ca=0.01$, $\alpha=1.7$.  
(b--e) Instability intervals (enclosed regions) under varying $Sc$, $Ca$, $Pe_\phi$, and $Ma$, respectively.}
  \label{fig:2}
\end{figure}

%  \caption{(a) The relations of $\omega_i$ on $k$ across different values of $\delta$. The gray-shaded area corresponds to $\omega_i > 0$, indicating instability. Stability is recovered upon the removal of the Marangoni force ('no Marangoni force'). The controling parameters are $Ma=10^4$, $Ca=0.01$, and $\alpha=1.7$. 
%  (b) The instability intervals (enclosed regions) under different physical parameters. Red curves indicate variations in $Ca$ number, blue curves correspond to variations in the $Ma$ number, and green curves depict variations in $\alpha$.
%  }

We identify an instability that arises solely in diffuse interfaces and is absent in the sharp-interface limit, driven by Marangoni stresses. We term this the \textit{diffuse-interface Marangoni instability}.

As shown in Figure~\ref{fig:2}a, short-wavelength instabilities emerge when the interface has finite thickness, even under conditions where sharp-interface theory predicts stability (equal viscosity and diffusivity in both phases)~\citep{sternling_interfacial_1959,schwarzenberger_pattern_2014}. For $Ma = 10^4$, $Ca = 0.01$, $\alpha = 1.7$, and $Sc=1000$, an unstable band appears at $\delta=0.2$ and narrows as $\delta$ decreases, disappearing entirely as $\delta$ approaches $0.02$—an estimated sharp-interface limit.
Disabling Marangoni stresses by omitting the $-\text{i}k\tilde{c} W^b$ term in Eq.~\ref{eq:linear_u}  suppresses the instability, yielding a single $\delta$-independent growth curve  (blue triangles in Fig.\ref{fig:2}a).
This confirms that the observed instability stems from solutal Marangoni effects rather than numerical artifacts.

To probe the instability mechanism, we vary key parameters individually: 
Marangoni number $Ma \in [10^3, 10^5]$, capillary number $Ca \in [10^{-3}, 10^{-1}]$, Schmidt number $Sc \in [10^2, 10^4]$, and mobility index $\alpha \in [1.3, 1.7]$. 
These span $Re \in [1, 100]$ and $Pe_\phi \in [0.3, 1.6\time 10^5]$, with $Re = Ma/Sc$ and $Pe_\phi = UL/\delta^{\alpha}$.
Default values are $Ma = 10^4$, $Ca= 0.01$, $\alpha = 1.7$, and $Sc=1000$.

Figure~\ref{fig:2}b-e show the neutral stability curves $k_{NS}$ versus interfacial thickness $\delta$. 
$Sc$ has little influence, consistent with prior studies~\citep{reichenbach_linear_1981,LI2025105073}. 
In contrast, $Ma$, $Ca$, and $Pe_\phi \sim \delta^{-\alpha}$ significantly alter the instability. 
Lower $Ca$ or higher $\alpha$ suppress short-wavelength modes but leave the critical thickness $\delta_{\mathrm{cr}}$ unchanged. 
In contrast, reducing $Ma$ narrows the unstable wavenumber band and increases $\delta_{\mathrm{cr}}$, shrinking the unstable regime. 
These results indicate that while $\delta$, $Ma$, $Ca$, and $Pe_\phi$ govern instability onset, $\delta_{\mathrm{cr}}$ is uniquely determined by $Ma$.

\subsection{Scaling laws for critical diffuse-interface thickness}
\label{subsec:3.2}

\begin{figure}
  \centerline{\includegraphics[width=\linewidth]{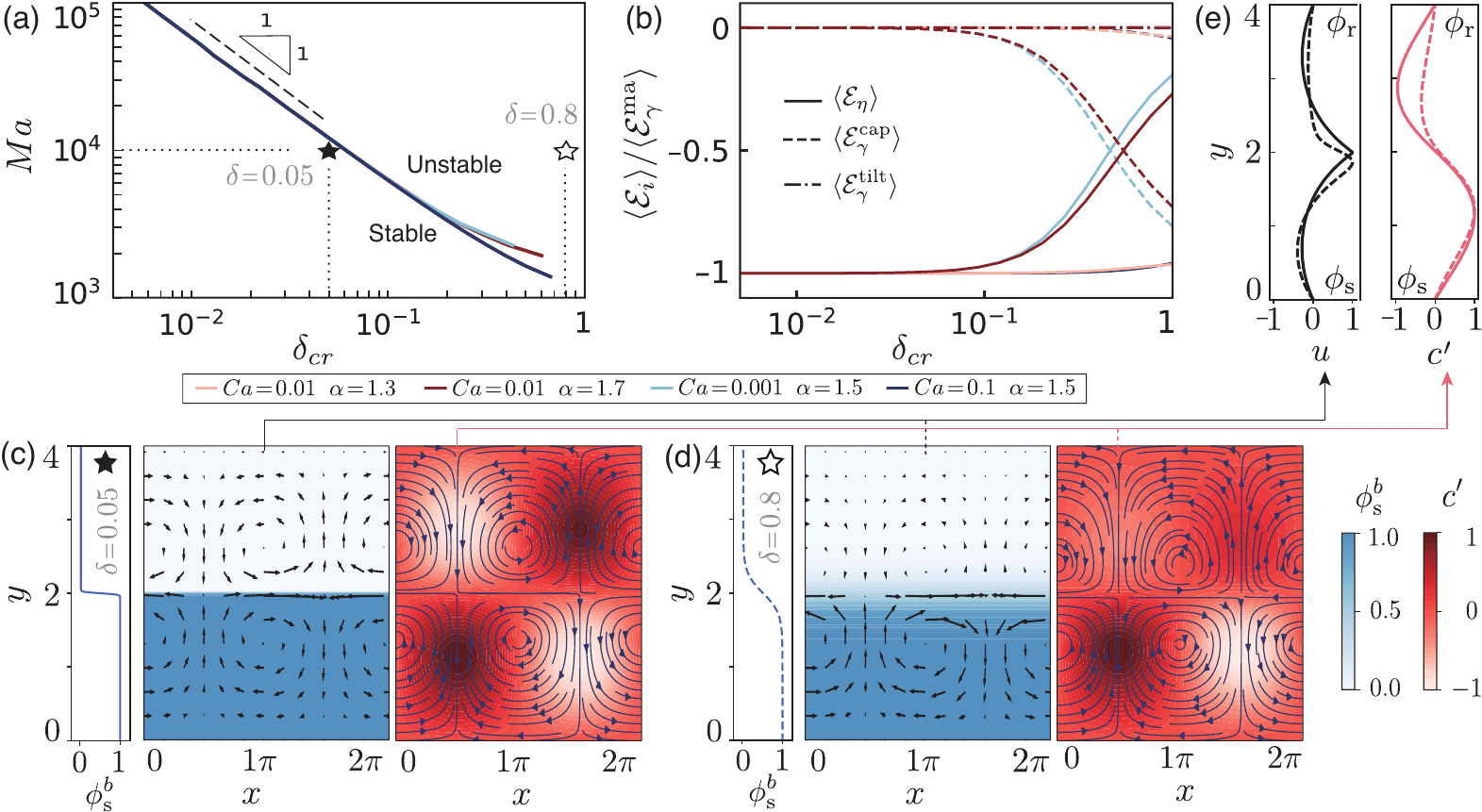}}
\caption{
(a) Scaling of the critical interface thickness $\delta_{\mathrm{cr}}$ with $Ma$, showing $\delta_{\mathrm{cr}} = 590/Ma$ for $Ma \gtrsim 4000$, independent of $Ca$ and $\alpha$. Deviations from the scaling emerge with increasing $\alpha$ or decreasing $Ca$.  
(b) Energy budget showing the relative contributions of viscous dissipation, capillary forces, and Marangoni stresses as $\delta_{\mathrm{cr}}$ increases.  
(c,d) Base-state phase field $\phi^b_\text{s}$ and solute perturbation $c^\prime$ for the stable case ($\delta = 0.05$) and unstable case ($\delta = 0.8$). Velocity vectors are overlaid on $\phi^b_\text{s}$; streamlines are shown in $c^\prime$.  
(e) Cross-sectional profiles of $u$ (at $x = \pi$) and $c^\prime$ (at $x = \pi/2$) along the $y$-direction for the stable (solid lines) and unstable (dashed lines) cases.  
}
  \label{fig:3}
\end{figure}

In the laminar regime, the critical thickness $\delta_\mathrm{cr}$ depends uniquely on $Ma$, not $Ca$ or $Pe_\phi$, indicating that $\delta_\mathrm{cr}$ is primarily governed by solutal Marangoni advection–diffusion dynamics near the interface (Eqns.~\ref{eq:c} and \ref{eq:linear_c}). 
The circulation (Fig.~\ref{fig:1}b) transports solute toward the interface, steepening concentration gradients, while diffusion tends to smooth them.
This competition features the Marangoni interfacial flows~\citep{manikantanSurfactantDynamicsHidden2020}.
Balancing diffusion (timescale $t_D\sim(\delta L)^2/D$) and advection ($t_A\sim \delta L /U$) yields the scaling, 
\begin{equation} 
  \label{eq:relationship}
\delta_{\mathrm{cr}}\sim Ma^{-1}.
  \end{equation} 
 As $Ma$ increases, stronger advection requires a thinner interface to maintain stability. 
 This $Ma$\,-\,$\delta_{\mathrm{cr}}$ scaling is confirmed as $\delta_\mathrm{cr} \rightarrow 0$ in Figure~\ref{fig:3}a, and remains robust under changes in $Ca$ or $Pe_\phi$, as shown in Figures~\ref{fig:2}cd.
This relation provides a physical criterion for selecting $\delta$ in diffuse-interface simulations with Marangoni effects, ensuring convergence to the sharp-interface limit~\cite{jacqmin_calculation_1999}. 
While thin-interface limit criteria exist for passive two-phase flows~\citep{magalettiSharpinterfaceLimitCahn2013,demont_numerical_2023}, the present finding extends them to Marangoni-driven instabilities, consistent with thermocapillary flow observations by~\citet{verschuerenDiffuseinterfaceModellingThermocapillary2001}.
 
To explain the deviation from the scaling law~(Eqn.~\ref{eq:relationship}) at larger interfacial thickness (i.e., for $Ma \lesssim 4000$), we perform an energy-based analysis to quantify the competing mechanisms.
Multiplying the linearized perturbation equations (Eqns.~\ref{eq:linear_u} and ~\ref{eq:linear_v}) by the velocity perturbations yields an energy budget,
\begin{equation}
  \label{eq:energy_budget2}
\begin{aligned}
   \frac{\partial E_\text{kin}}{\partial t} = &\underbrace{\frac{{1}}{{Re}}   \left[-\left(\nabla u^{\prime}\right)^2-\left(\nabla v^{\prime}\right)^2\right] }_{\mathcal{E}_\eta:\, \mathrm{Viscous}~\mathrm{dissipation}} + \underbrace{\frac{{1}}{{Re}}\frac{1}{Ca}  \left[ \frac{\Gamma^b}{\left|\nabla \phi_\text{s}^b\right|} \frac{\partial^2 \phi_\text{s}^{\prime}}{\partial x^2}  v^{\prime}  \right] W^b}_{\mathcal{E}^\mathrm{cap}_\gamma:\, \mathrm{Capillary}~\mathrm{effect}} \\
  &+ \underbrace{\frac{{1}}{{Re}}\frac{1}{Ca}\left[\frac{1}{\left|\nabla \phi_\text{s}^b\right|} \frac{\partial \phi_\text{s}^{\prime}}{\partial x} \frac{\mathrm{d} \Gamma^b}{\mathrm{d} y} u^{\prime}\right] W^b  }_{\mathcal{E}^\mathrm{tilt}_\gamma:\, \mathrm{Tilt}-\mathrm{induced}~\mathrm{Marangoni}}+ \underbrace{ \frac{{1}}{{Re}}\left[-\frac{\partial c^{\prime}}{\partial x} u^{\prime}\right] W^b }_{\mathcal{E}^\mathrm{mag}_\gamma:\, \mathrm{Solutal}~\mathrm{Marangoni}~\mathrm{effect}}.
\end{aligned}
\end{equation}
Here, the kinetic energy density is defined as $E_\text{kin}=\rho^b ({u^{\prime}}^2+{v^{\prime}}^2)/2$, since the base flow $\boldsymbol{U}^b$ is zero and total velocity reduces to the perturbation field   ($\boldsymbol{u}=\boldsymbol{u}^\prime$).
The prefactor $Ma/Sc$ is replaced by $Re\lesssim 4$.
%The derivation is given in Appendix~\ref{sec:energy}.

The viscous dissipation $\mathcal{E}_\eta$ reflects energy loss due to shear.
The solutal Marangoni term $\mathcal{E}_\gamma^\mathrm{mag}$, originates from tangential solute gradients $-\partial c^\prime/\partial x$.
The capillary term $\mathcal{E}_\gamma^\mathrm{cap}$ arises from interface curvature $\partial^2 \phi^\prime_\text{u}/\partial x^2$ and base-state tension $\Gamma^b$.
The tilt-induced Marangoni term $\mathcal{E}_\gamma^\mathrm{tilt}$ stems from interfacial tilt $\partial \phi_\text{u}^{\prime}/\partial x$, which projects normal interfacial tension gradients $\mathrm{d} \Gamma^b/\mathrm{d} y$ across the diffuse layer~\citep{berry1971molecular,marchand2011surface} onto the tangential direction -- an effect inherently tied to the finite thickness of diffuse interfaces and absent in sharp-interface models.

Along the $Ma$\,-\,$\delta_\mathrm{cr}$ paths (Fig.~\ref{fig:3}a) , integrating over the domain $[2\pi/k \times 4L]$ gives the power balance $\langle\mathcal{E}_\eta\rangle+\langle\mathcal{E}_\gamma^\mathrm{cap}\rangle+\langle\mathcal{E}_\gamma^\mathrm{tilt}\rangle+\langle\mathcal{E}_\gamma^\mathrm{mag}\rangle=0$ ($\omega_i=0$).
Divergence terms vanish under periodic, homogeneous boundary conditions~\citep{zhang_linear_2013}.
Figure~\ref{fig:3}b shows the normalized contributions of each term by $\langle\mathcal{E}_\gamma^\mathrm{mag}\rangle$ as $\delta_\mathrm{cr}$ increases. 
For small $\delta_\mathrm{cr}$ ($\lesssim 0.1$), the balance is primarily between viscous dissipation $\mathcal{E}_\eta$ and solutal Marangoni forcing $\mathcal{E}_\gamma^\mathrm{mag}$, consistent with the scaling law. 
As $\delta_\mathrm{cr}$ increases, the influence of capillary $\mathcal{E}_\gamma^\mathrm{cap}$ and tilt-induced Marangoni effects $\mathcal{E}_\gamma^\mathrm{tilt}$ becomes significant at a certain $\delta_\mathrm{cr}$, where the deviation from the ideal $\delta_\mathrm{cr}\sim Ma^{-1}$ relation starts exactly. 
Thus, interface deformation and non-uniform tension projection across the diffuse layer play key roles in setting the limits of the scaling law's validity.

Figures~\ref{fig:3}cd show the base-state phase filed $\phi^b_\text{s}$ (blue), velocity field $\boldsymbol{u}=\boldsymbol{u}^\prime$ (arrows), and perturbation concentration field $c^\prime$ (red) for two representative cases: a stable system ($\delta = 0.05$) and an unstable one ($\delta = 0.8$), corresponding to the solid and open star symbols in Figure~\ref{fig:3}a, respectively.
In the stable case (Fig.~\ref{fig:3}c), vortices are symmetrically distributed about the interface, as evidenced by the velocity component $u$ profiles at $x=\pi$ (Fig.~\ref{fig:3}e, black solid line).
This balanced Marangoni recirculation gives rise to an anti-symmetric $c^\prime$ field, confirmed by the profile at $x=\pi/2$ (Fig.~\ref{fig:3}e, red solid line). 
In contrast, the unstable case (Fig.~\ref{fig:3}d) exhibits anti-symmetry breaking, with flow field shifting downward into source phase (Fig.~\ref{fig:3}e, black dashed line).
The zero-crossing point of $c^\prime$ shifts upward from the interface centerline ($y=2$), indicating solute enrichment in the receiver phase and depletion in the source phase (Fig.~\ref{fig:3}e red dashed line). 
These symmetry-breaking shifts in $u$ and $c'$ motivate the next section, in which we investigate their physical origin and impact on the instability.

\subsection{Mechanism of instability in diffuse interfaces: Anti-flux vs. Pro-flux shifts}
\label{subsection:3.3}

\begin{figure}
  \centerline{\includegraphics[width=\linewidth]{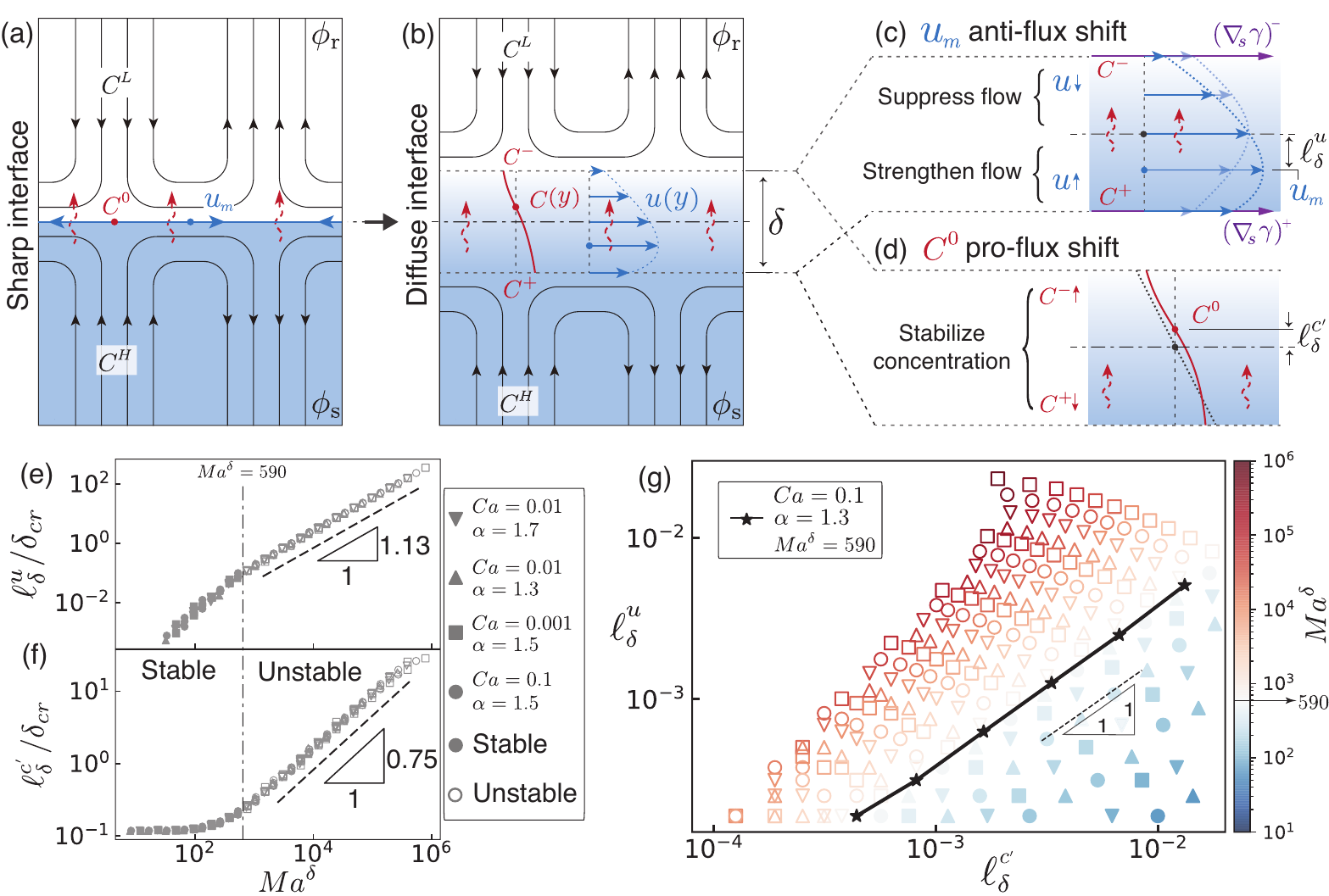}}
  \caption{
    (a) Schematic of interfacial solute and velocity fields in the sharp-interface limit, where symmetric advection yields no net concentration change. 
    (b) Profiles of solute concentration $C(y)$ and tangential velocity $u(y)$ across the diffuse interface.% with interfacial gradients spread over a finite thickness. 
    (c) Mechanism of velocity offset $\ell_\delta^u$: asymmetric solute advection enhances Marangoni flow on the source side, displacing the velocity peak away from the centerline.
    (d) Mechanism of solute offset $\ell_\delta^{c^\prime}$: coupled vertical flow and interfacial diffusion shift the position of zero net solute change.
    (e, f) Normalized offsets $\ell_\delta^u/\delta_{cr}$ and  $\ell_\delta^{c^\prime}/\delta_{cr}$ versus the modified Marangoni number $Ma^\delta = Ma \, \delta$, showing collapse across varying $Ca$ and $\alpha$. Instability sets in at $Ma_\text{cr}^{\delta}\approx 590$.
      (g) Phase diagram in the $\ell_\delta^u$\,--\,$\ell_\delta^{c^\prime}$ plane. 
      Stable (solid) and unstable (open) regimes are separated by the scaling $\ell_\delta^u \sim (\ell_\delta^{c^\prime})$.
}
  \label{fig:4}
\end{figure}

In classical sharp-interface systems with equal viscosity and diffusivity between two fluids, interfacial Marangoni instability is absent~\citep{sternling_interfacial_1959}. 
As illustrated in Figure~\ref{fig:4}a, symmetric convection brings high ($C^H$) and low ($C^L$) concentrations toward the interfacial stagnation point.
Equal diffusivity and viscosity ensure that net interfacial concentration remains unchanged, yielding no amplification of the Marangoni flow $u_m$.
In contrast, a diffuse interface introduces local asymmetry.
Figure~\ref{fig:4}b shows that inflow of low-concentration fluid reduces interfacial concentration $C^-$ on the receiver phase $\phi_\text{r}$ side, while inflow of high-concentration fluid increases $C^+$ on the source $\phi_\text{s}$ side.
Within the interfacial layer of thickness $\delta$, the resulting imbalance cannot be instantly homogenized, producing a tangential concentration gradient $C(y)$ and velocity profile $u(y)$.

As shown in figure~\ref{fig:4}c, the local concentration decrease on the receiver side ($\phi_\text{r}$)  reduces the interfacial tension gradient $(\nabla_s \gamma)^-$, thereby weakening the Marangoni flow~\citep{anderson1998diffuse}.
Conversely, the increased concentration on the source side ($\phi_\text{s}$) enhances $(\nabla_s \gamma)^+$, strengthening the flow.
This asymmetry in surface tension gradients induces a shear layer within the interface, with suppressed velocity near $\phi_\text{r}$ and enhanced velocity near $\phi_\text{s}$.
 The resulting shear promotes an interfacial instability reminiscent of the Kelvin–Helmholtz type~\citep{chandrasekhar2013hydrodynamic}. 
 We quantify the shift of the peak velocity ($u_m$) away from the interface centerline ($y=2L$) by the anti-flux offset, $\ell_{\delta}^{u}=2L-y_{u_m}$, which characterizes the destabilizing effect of advection-driven shear.
Diffusion counters this destabilization by smoothing concentration gradients ($C^-\uparrow$ and $C^+\downarrow$), as depicted in Figure~\ref{fig:4}d.
The position where the net concentration change vanishes ($C^0$) shifts along the solute flux into the receiver side. 
This stabilizing effect is captured by the pro-flux offset $\ell_{\delta}^{c^\prime}=y_{C^0}-2L$.

To quantify the competition between advection and diffusion across the interface, we introduce a modified Marangoni number $Ma^\delta=Ma\, \delta$, using the interfacial thickness $\hat{\delta}$ as the characteristic length scale.
As shown in Figures~\ref{fig:4}ef, both the velocity offsets $\ell_{\delta}^{u}$ and the concentration offset $\ell_{\delta}^{c^\prime}$ increase monotonically with $Ma^\delta$,  following distinct power-law trends that collapse across different $Ca\in [10^{-3}, 10^{-1}]$ and $\alpha \in [1.3, 1.7]$, confirming their universality.
A critical transition emerges at $Ma^\delta \approx 590$: 
below this threshold, the system remain stable, with $\ell_{\delta}^{u}$ growing sublinearly and $\ell_{\delta}^{c^\prime}$ superlinearly; 
Beyond it, the system becomes unstable, and both offsets grow with fixed exponents, approximately 1.13 for $\ell_{\delta}^{u}$ and 0.75 for $\ell_{\delta}^{c^\prime}$.

To further capture this interplay,  Figure~\ref{fig:4}g plots $\ell_{\delta}^{u}$ against $\ell_{\delta}^{c^\prime}$, revealing a clear separation between stable (bluish solid) and unstable (reddish open) regimes by the power-law relation  $\ell_{\delta}^{u} \sim (\ell_{\delta}^{c^\prime})^m$.
The critical exponent $m=1$ at $Ma^\delta=590$ (black solid line with star symbols) marks a crossover from diffusion-dominated stabilization ($m<1$) to advective-driven destabilization ($m>1$), highlighting the offset competition as the underlying mechanism governing interfacial instability.

\section{Conclusion}
\label{sec:4}

We have uncovered a Marangoni-induced diffuse-interface instability that is absent in the sharp-interface limit. 
Through linearized analysis of a validated phase-field NSAC model, we show that finite interfacial thickness introduces asymmetric solute distributions and tangential velocity shifts that destabilize the interface, even in fluids with matched viscosity and diffusivity. 
A critical interfacial thickness $\delta_\text{cr} \sim Ma^{-1}$ emerges from the competition between diffusive and advective timescales. 
Energy analysis reveals that deviations at large $\delta_\text{cr}$ arise from interface deformation and asymmetric tension projection.
By introducing a modified Marangoni number $Ma^\delta = Ma \delta$, we reveal universal power-law behavior in velocity and concentration offsets, independent of $Ca$ and $\alpha$. 
The crossover at $Ma^\delta \approx 590$, where the exponent $\ell_{\delta}^{u} \sim (\ell_{\delta}^{c^\prime})^m$ with $m=1$, marks the transition from diffusive stabilization to advective destabilization.

These findings not only clarify the physical origin of instability in diffuse interfaces but also provide a predictive framework for multicomponent multiphase systems where sharp-interface assumptions fail. 
This work opens new directions for studying interfacial transport in miscible fluids, soft materials, and microfluidic systems~\citep{shim2022diffusiophoresis,michelin2023self,ault2024physicochemical}, where interfacial thickness and multi-physics coupling critically influence dynamics.

H.T. acknowledges support from the National Natural Science Foundation of China (Grant Nos. 12472271, 12102171) and the Guangdong Basic and Applied Basic Research Foundation (Grant No. 2024A1515010614). M.Z. acknowledges support from the Ministry of Education, Singapore (WBS No. A-8001172-00-00).

Declaration of Interests: The authors report no conflict of interest.

\bibliographystyle{jfm}

\bibliography{jfm-instructions}

\end{document}